*Analysis, Design and Fabrication of centimeter-wave Dielectric Fresnel Zone Plate Lens and reflector*

A. Mahmoudi , Physics group, University of Qom, Iran

Abstract: Fresnel lens has a long history in optics. This concept at non-optical wavelengths is also applicable. In this paper we report design and fabrication of a half and quarter wave dielectric Fresnel lens made of Plexiglas, and a Fresnel reflector at 11.1 GHz frequency. We made two lenses and one reflector at same frequency and compare their gain and radiation pattern to simulated results. Some methods for better focusing action will be introduced.

I. Introduction

Fresnel Zone Plate (FZP) is a planar structure that can convert an incident plane wave into a spherical wave front that converges at a focal point. Depending on type of material of FZP, we have FZP reflector (metal) or FZP lens (dielectric). In both types, this is diffraction not refraction that causes convergence of spherical reflected/transmitted wave. A circular FZP consists of a number of circular Fresnel zones that their radii are determined by:

$$r_n = \sqrt{(F + \frac{n\lambda_0}{p})^2 - F^2} \quad n=0,1,2,\ldots,N \quad (1)$$

Where F is focal length, $\lambda_0$ is free-space wavelength and p is an integer. Circular ring between two adjacent circles is a Fresnel zone. If p=1, Fresnel zones will be called full-wave zones and if p=2 these zones are half-wave zones and so on. In simple FZPs , odd/even zones are opaque/absorbing. Reflection form surface of such metallic Fresnel zones or transmission through a transparent dielectric Fresnel zones will result in amplitude amplification. Based on scalar diffraction theory and using geometry shown in Fig.1 we can calculate distribution of field amplitude in focal plane as [1]:

$$E(\theta) = \frac{2\pi(1+\cos\theta)e^{j(\omega t - kR)}}{R} e^{j\delta n}$$
$$\cdot [\int_0^{r_n} F(\rho)J_0(k\rho\sin\theta)\rho d\rho - \int_0^{r_{n-1}} F(\rho)J_0(k\rho\sin\theta)\rho d\rho] \quad (2)$$

Where exponent factor $e^{-j\delta n}$ denote phase shift applied by plate material on incident wave , $r_n$ and $r_{n-1}$ are radii of n-th and n-1-th zones respectively , $F(\rho)$ is a function that indicates incident field distribution on plate and $J_0$ is first kind zero order Bessel function. So the far-field pattern problem can be reduced to solving the integral:

$$I(u) = \int_0^a F(\rho)J_0(u\rho)\rho d\rho \quad (3)$$

Equation (3) cannot be calculated analytically and we must evaluate it numerically. Making this, we have field distribution as a function of $\theta$.

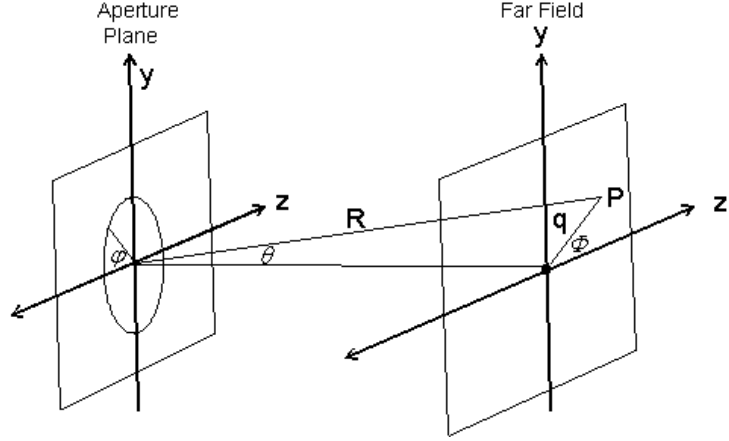

Fig.1: geometry for calculating far field distribution

II. Modified FZP's

In simple FZP's, it is only a fraction (half) of plate surface that contributes in concentrating wavefront. In such conditions we need a plate with large area to increasing gain to a sufficient level. If the last zone have radius $r_N$ is equal to radius of the plate (D/2), solving equation (1) for N we have

$$N = \frac{2D}{\lambda_0}\left[\sqrt{(\frac{F}{D})^2 + \frac{1}{4}} - (\frac{F}{D})\right] \qquad (4)$$

From above equation, it is apparent that for constant (F/D), N will increase directly with $D/\lambda_0$. In optical wavelengths and lower (e.g. x-ray), there will be a large number of Fresnel zones on a plate with a small diameter and without any modification we will have high gain. But at centimeter wavelengths we will need large surface simple FZP's. Increasing FZP diameter will increase weight and cost of lens or reflector. Thus we must increase number of Fresnel Zones without diameter increasing. To utilize the opaque zone apertures, Wiltse [2] replaced the reflecting (or absorbing) rings by phase-reversing (half-wave) dielectric rings. Based on this lens various FZP antennas with a radiation efficiency of 25-30% were developed and studied [2], [3]. In some other works [4], this phase correction is caused by using rings with different $\varepsilon_r$. By dividing each full-wave Fresnel zone to a number of subzones and making appropriate changes in them (e.g. changing depth by cutting groves) we will expect better results. By a ray tracing analysis using geometry shown in Fig.3, we can drive appropriate depth of phase step (d) necessary for a $\pi$ phase shift:

$$d = \frac{\lambda_0}{2(\sqrt{\varepsilon_r} - 1)} \qquad (5)$$

Where $\varepsilon_r$ is the relative permeability of FZP lens material and d is phase step depth. This is called phase reversing zone plate. If a plane wave is normally incident on zone plate, the portions of radiation which pass through or reflect from various parts of the transparent (reflecting) zones all reach the selected focal point with phases which differ by less than one-half period. Thus, the zone plate acts like a lens, producing a focusing action on the radiation it transmits or reflects. In case of dividing each full zone to m phase step (subzone) Eq.(5) will have this form:

$$d = \frac{\lambda_0}{m\left(\sqrt{\varepsilon_r} - 1\right)} \qquad (6)$$

Here the depth of s-th step is equal to sd ($S \leq m$).

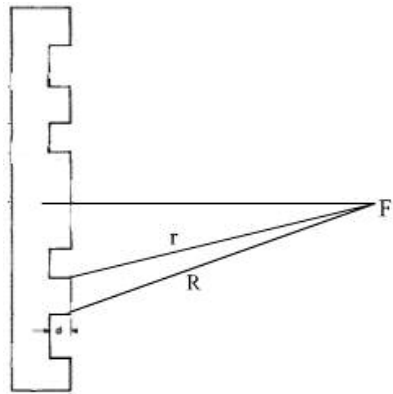

Fig.2: phase step in Fresnel lens/reflector

III. Calculations

Based on equation (2) and using a Matlab code for numerical integration, we can calculate field distribution around focal point. Equation (2) gives n-th zone contribution in total field amplitude at the focal point. If we divide each full-wave zone on dielectric FZP lens into m suzones and cutting groves with depths resulted from Eq. (6), then the total field amplitude at the focal point is equal to this weighted sum:

$$A_{Total} = \sum_{n=1}^{n=N} \sum_{S=0}^{S=m-1} e^{\frac{j2\pi S}{m}} A(n,S) \qquad (7)$$

Where $e^{\frac{j2\pi S}{m}}$ is phase shift related to S-th subzone and A(n,S) is S-th subzone contribution, and N is total number of full zones on the FZP. Using a Matlab code , radiation pattern for half-wave (m=2) and quarter-wave(m=4) FZP lens were calculated , results are shown in Fig.3 and Fig.4.

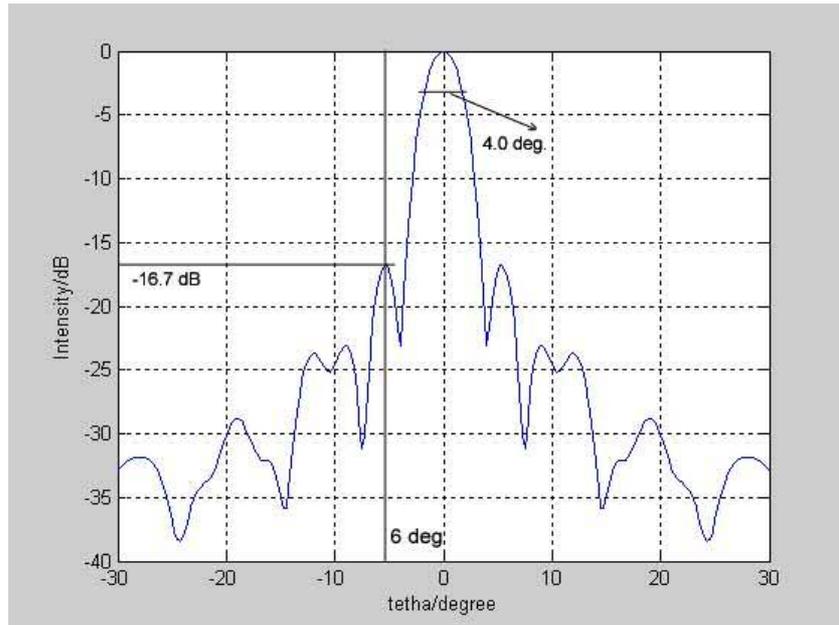

Fig.3: calculated radiation pattern (E-plane) for 2 step Fresnel lens.
D=40cm and $\lambda = 2.7$ cm

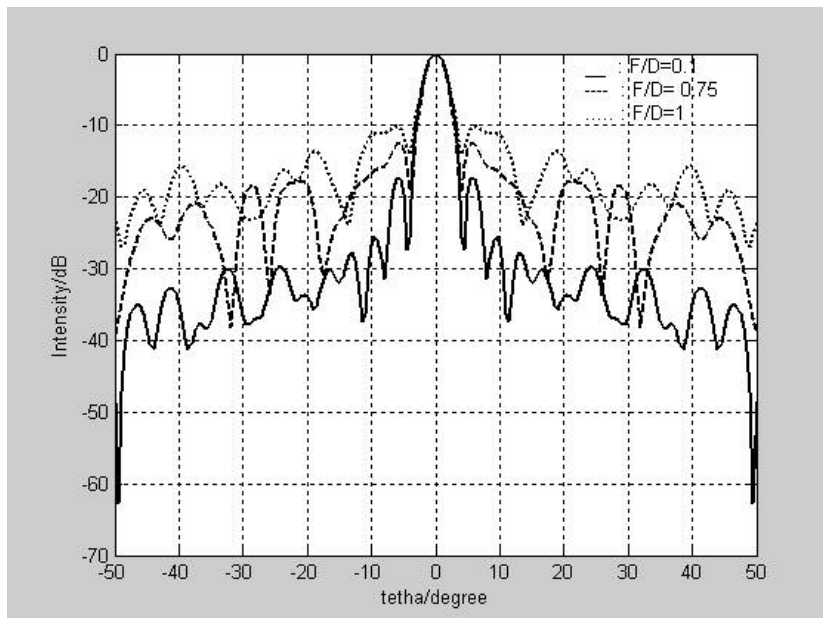

Fig.4: calculated radiation pattern for 4-step Fresnel lens
For different F/D, D=40cm and $\lambda = 2.7$ cm

IV. Experimental

We used Plexiglas (Polymetthylmethacrylate) with mass density of 1.19 g/cm$^3$ as Lens material. Wavelength $\lambda$ is 2.7cm. For $\varepsilon_r$ measurement we applied two techniques, a Pseudo-Brewster's Angle Method [5] and a waveguide method, from both techniques it found to be 2.63. First we fabricated a Fresnel lens with m=2 (Fig.5 & Fig.6). Design parameters of this lens are listed in table.1.

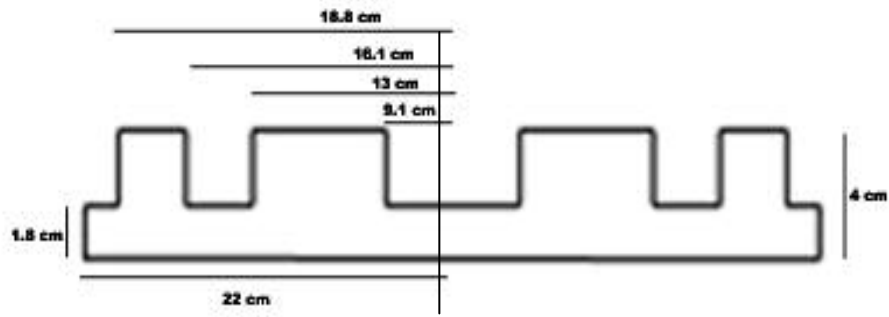

Fig.5: Dimensions of the fabricated Lens

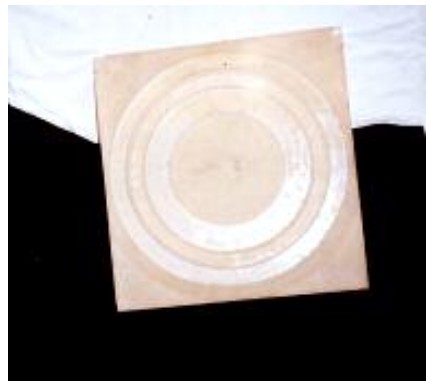

Fig.6: Fabricated 2-step Fresnel lens

Table.1: design parameters for fabricated 2-step FZP lens

| Diameter (cm) | 40 |
|---|---|
| F/D | 0.75 |
| N (number of full zones) | 2 |
| m (number of subzones on each full zone) | 2 |
| Thickness (cm) | 4 |

Then by cutting, we converted each zone into a concave surface, then covered it by Aluminum sheet, this is a segmented Fresnel reflector (Fig.7).

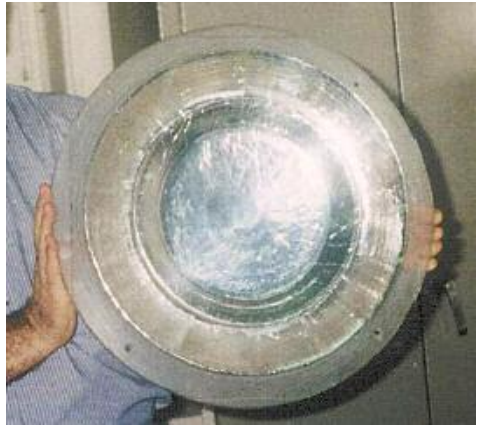
Figure.7: segmented Fresnel lens

Using a horn antenna, radiation pattern of fabricated lens and reflector were measured (Fig.8 & Fig.9)

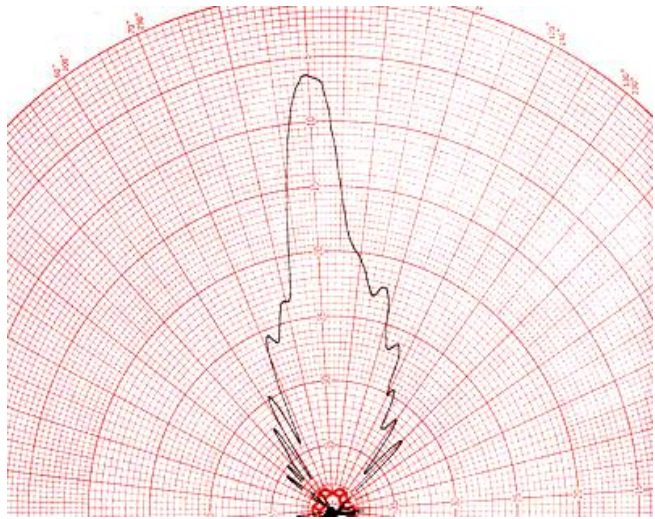

Fig.8: E-plane radiation pattern for 2-step lens
Freq. =11.1GHz.

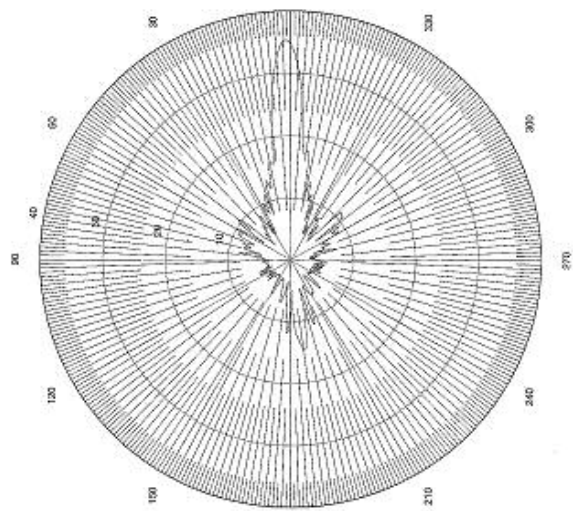

Fig.9 : Measured radiation pattern for fabricated 4-step dielectric lens.
Freq. =11.1 GHz

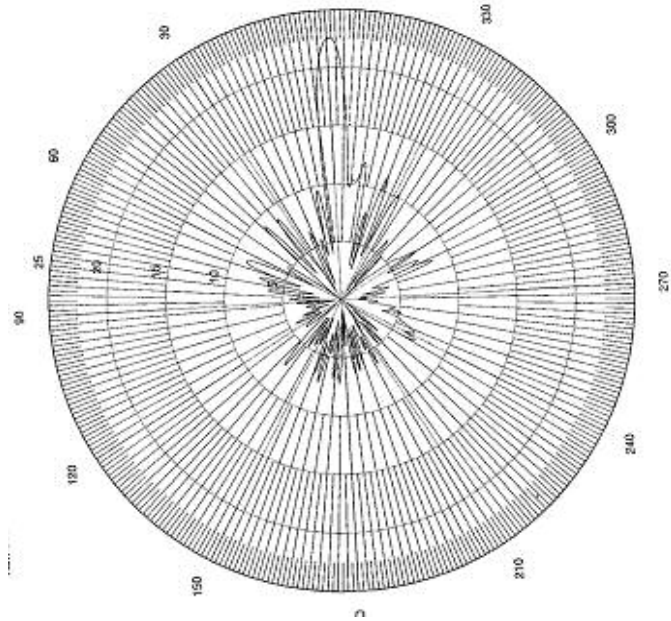

Fig.9:E-plane radiation pattern for segmented continuous Fresnel reflector (shown in fig2.) ,Freq.= 11.1GHz

Then we made another lens with m=4. an schematic view of this lens is shown in Fig.10 and Fig.11.

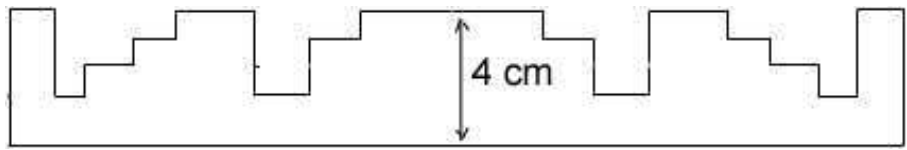

Fig.10: side view of lens with m=4

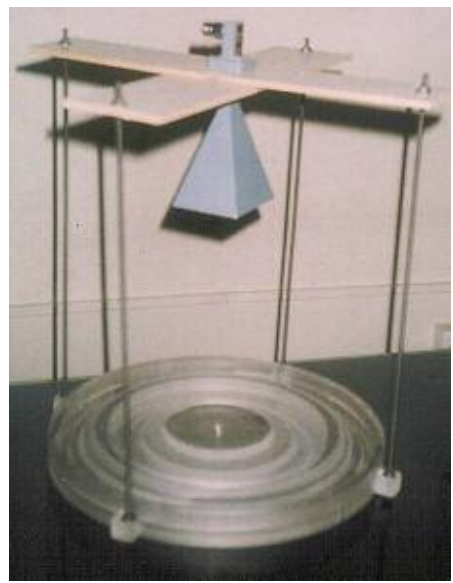

Fig.11: Fabricated quarter-wave lens (m=4) with horn feed.

V. Conclusions

Radiation pattern and gain of fabricated lenses and reflector were made in K.N.T University of technology, Tehran, Iran. In these measurements, we used a horn feed with gain equal to 17.5 dB. Radiation patterns are shown in Fig.8, Fig.9 and Fig.12. Other measurements are shown if Table.2. These results are in good agreement with calculated radiation patterns shown in Fig.3 and Fig.4. For example, from Fig.3, calculated first sidelobe level is -16.7dB and measured value is -16.dB (Fig.8), also 3dB width of measured radiation pattern (Fig.8) and calculated radiation pattern (Fig.3) are very close to each other. The measured value is 5 degree and calculated value is 4 degree.

Table.2: Measured radiation pattern properties of fabricated items

| system | Plane | 3dB width (deg.) | First sidelobe angle (deg.) | First sidelobe Level (dB) | Gain (dB) |
|---|---|---|---|---|---|
| 2-step lens +horn feed | H | 5.5 | 11 | -16 | 29 |
|  | E | 5.5 |  |  |  |
| 4-step lens + horn feed | H | 5.5 | 15 | -22 | 28 |
|  | E | 5 | 10.5 | -15.7 |  |
| Segmented reflector + horn feed | H | 7 | 10 | -10.3 | 19.6 |
|  | E | 6 | 9 | -13.4 |  |

As can be seen from Table.2, segmented reflector has the worst results; it can be caused by shadowing effect due to the wooden fixtures applied for horn feed mounting (Fig.11). Comparison between 2-step and 4-step lenses reveals that they have nearly same 3dB width, but the 4-step lens has a sharper pattern than 2-step lens, the 2-step lens has higher gain. This is because the greater thickness of 4-step lens (Fig.10) in central region in comparison to 2-step lens (Fig.5).

VI. Acknowledgement

The author wishes to thank R. Afzalzadeh and S. M. Aboutorab for their help.